\documentclass[10pt, conference, compsocconf]{IEEEtran}
\usepackage{graphicx}
\usepackage{color}
\usepackage{url}
\usepackage{hyperref}
\usepackage{multirow}
\usepackage{amsmath}
\begin{document}

%
\title{Mining Illegal Insider Trading of Stocks:  A  Proactive Approach}

\author{\IEEEauthorblockN{Sheikh Rabiul Islam}
\IEEEauthorblockA{Computer Science\\
Tennessee Technological University\\
Cookeville, U.S.\\
sislam42@students.tntech.edu}
\and
\IEEEauthorblockN{Sheikh Khaled Ghafoor}
\IEEEauthorblockA{Computer Science\\
Tennessee Technological University\\
Cookeville, U.S.\\
sghafoor@tntech.edu}
\and
\IEEEauthorblockN{William Eberle}
\IEEEauthorblockA{Computer Science\\
Tennessee Technological University\\
Cookeville, U.S.\\
weberle@tntech.edu}
}

\maketitle

\begin{abstract}
Illegal insider trading of stocks is based on releasing non-public information (e.g., new product launch, quarterly financial report, acquisition or merger plan) before the information is made public. Detecting illegal insider trading is difficult due to the complex, nonlinear, and non-stationary nature of the stock market. In this work, we present an approach that detects and predicts illegal insider trading proactively from large heterogeneous sources of structured and unstructured data using a deep-learning based approach combined with discrete signal processing on the time series data. In addition, we use a tree-based approach that visualizes events and actions to aid analysts in their understanding of large amounts of unstructured data. Using existing data, we have discovered that our approach has a good success rate in detecting illegal insider trading patterns.  
\end{abstract}
\begin{IEEEkeywords}
 Stock Market; Neural Network; Natural Language Processing; Time Series Prediction; Illegal Insider Trading;
\end{IEEEkeywords}

\IEEEpeerreviewmaketitle

\section{Introduction}\label{introduction}
The stock market is a collection of markets and exchanges where equities, bonds and other form of securities are issued and traded. It provides companies access to capital in exchange for a slice of ownership with the investors.  The largest stock exchange in the world is the New York Stock Exchange (NYSE). NASDAQ is the second largest in the U.S., where most of the tech companies participate. The Security and Exchange Commission (SEC) is the regulatory body that oversees the U.S. stock market.  

Securities fraud are deceptive activities in connection with the offer and sale of the securities \cite{golmohammadi2014detecting}. According to the Federal Bureau of Investigation (FBI) report \cite{fbi_2010}, the following are the most prevalent types of fraud being encountered today in the securities market: 1) Market Manipulation\textemdash creating an artificial buying pressure for some usually low trading stock which is largely controlled by the perpetrators. This results in an illicit gain to the perpetrators and losses to the innocent and naive investors. Market manipulation destroys the fair and orderly market, 2) Late Day Trading\textemdash illicit purchase or sale of securities after regular market hours. This type of trading is restricted due to important market influencing  decisions are released after the close of regular trading, 3)	High Yield Investment Fraud\textemdash the offering of low or no risk investments that guarantee unusually high rates of return.

Insider trading in the stock market is trading based on non-public information. Insider trading in the U.S. is a profitable activity \cite{jeng2003estimating}. It can be both legal and illegal. A legal insider trading needs to be done by following the proper guideline of the regulatory. On the other hand, illegal insider trading occurs when trading is performed based on non-public (private, leaked, tipped) information (e.g., new product launch, quarterly financial status, acquisition or merger plan) before the information is made public.

Illegal insider trading detection and correct prediction of stock price are challenging problems. The traditional techniques do not adequately address these problems.  New approaches and techniques are required in order to understand the complexity of a stock market. For instance, Artificial Neural Network (ANN), an approach that is loosely modeled after the neural structure of a mammalian cerebral cortex can be useful in this regards. Neural Networks are organized into layers of interconnected nodes. Patterns are represented by a neural network via the input layers, which are connected to one or more hidden layers. These layers end up with the output layer where the result from the input pattern is produced. However, regular neural networks do not work well for time series or sequential data because they lack  memory (i.e., storage) of past events. As a result, a more popular neural network technique is a Recurrent Neural Network (RNN), where the output of a neuron is fed back to the input of the neuron again. However, while that gives some memory (limited and consecutive) to the neuron, it still suffers from the vanishing/exploding gradient problem. Furthermore, the learning rate of the network drops quickly as gradient contribution from far away steps become zero.

On the other hand, Long Short-Term Memory Networks (LSTM) overcome some of the problems of the RNN. Unlike the feed-forward network, LSTM can keep any number of the previous steps (as required) in the memory and forget those when needed. It also helps to manage the memory in a more controlled way\textemdash which information to keep, which to update, which to forget, and which to pay attention to over a longer period of time. And this is why LSTM is very useful in time series prediction where the data might have seasonality (cycle of behavior over time), noise (variability in data that can't be explained correctly with model) and trends (increasing or decreasing behavior over a time period).

\textbf{Contribution:} Our primary contribution in this work is the use of an LSTM RNN to predict stock volume for some targeted companies and then using our proposed algorithm ANOMALOUS, we will be able to detect  anomalous pattern in the predicted data that may be an effect of an illegal insider trading. Other contributions are in the data preprocessing stage: A) classification of illegal insider trading cases from a huge number of litigation related press releases (SEC litigation archives); B) implementation of a tree-based approach that visualizes events and actions to aid analysts in their understanding of large amounts of unstructured data (e.g., litigation related press releases). Our experimental results support that (1) it is possible to detect illegal insider trading within a short period of time from a mix of affected and non-affected companies, and (2) it is possible to classify and intuitively visualize a large number of unstructured data. Usually, investors become aware of illegal insider trading when the SEC files cases\textemdash too late for any actionable intelligence. Thus, what we are proposing is a computational intelligence-based approach that analyzes data from heterogeneous sources and detects potential illegal insider trading activities long before the official news of that information.  To the best of our knowledge, this is the first comprehensive attempt to mine illegal insider trading from a diverse source of data using computational intelligence, and as such, limits our ability to comprehensively comparing our proposed approach with other related work. In the following sections, we present previous research on predicting stock market manipulation. We then describe the dataset we used, including the data preprocessing.  We then include our experiments, results, and analysis, followed by some concluding observations and future work.

\section{Background}\label{background}
In the work of \cite{golmohammadi2015time}, the authors propose a Contextual Anomaly Detection Approach (CAD) for complex time series data of stock market. The anomaly detection approach is treated here as a contextual or local outlier detection approach based on the information that similar companies tend to show similar behavior irrespective of the time of the year. In this research, five groups of anomaly detection methods are studied as follows: window based, proximity-based, prediction based, Hidden Markov Model-based, and segmentation based. All of them have their own advantages and disadvantages. According to the author, none of them is a perfect candidate for this type of time series based anomaly detection. Rather they propose a Contextual Anomaly Detection approach which they found to outperform other state of art approaches such as KNN and Random Walk. The LSTM RNN  that we are using in this work uses day-based, window-based and entire-history based prediction.

Golmohammadi  et al.  \cite{golmohammadi2014detecting} mention that existing fraud detection approaches heavily rely on a set of rules based on expert knowledge. However, the current stock market is more dynamic in nature, where perpetrators are constantly devising new schemes, so there is need of scalable machine learning algorithms to identify market manipulation activities. They applied different types of machine learning algorithms like Naive Bayes, SVM, CART, KNN, and Random Forest. Empirical results show that Naive Bayes outperforms other learning methods in terms of the F2 measure. The work of \cite{golmohammadi2014detecting} is actually an extension of the work of Diaz et al. \cite{diaz2011analysis}. According to the authors, three techniques contribute heavily to market manipulation:
1)	Buying or selling stock at the end of the day or last quarter to affect the closing price, 2) Wash trades (pre-arranged trades) that will be reversed later or that has no actual risk to the seller or the buyer, and 3) Cornering the market by obtaining a major portion of stock. The dataset used in this work is the same one as used in the work of Diaz et al. \cite{diaz2011analysis}. This dataset contains the market manipulation cases reported by the SEC between January and December of 2003, which includes 31 dissimilar stocks, 8 manipulated stocks and 25 stocks similar to the manipulated stocks. The dataset we use in our work contains data from the beginning year (1996) of the litigation archive to the year 2018, and our focus is narrowed to predicting illegal insider trading\textemdash which is one of the more difficult types of stock market fraud to detect and predict.

According to \cite{allen1992stock}, stock market abuse can happen in three primary ways: 1) Information-based manipulation, where financial rumor is released to affect the price; 2) Action-based manipulation, where equity supply/demand is squeezed to do the manipulation; and 3) Trade-based manipulation, where the manipulation is done simply by buying and selling, which makes it difficult to categorize as illegal. Furthermore, illegal insider trading mostly is done through buying and selling which makes it difficult to detect or predict.

Moreover, price manipulation cannot be detected from an event of a single action. Instead, most of the time it consists of a series of actions and manipulation strategies evolving over the time, where manipulation behaviors are not obvious when mixed with a massive set of normal records. According to \cite{cao2015adaptive} and \cite{cumming2011exchange} price manipulation can be accomplished in many different forms.  For instance, there is ramping or gouging or momentum ignition, where the investor enters a buy/sell order (sometimes called  a ``spoofing order'') at a price which is much higher/lower than the actual bidding or asking price, creating a false appearance of interest followed by a ``bona fide'' order which is opposite of the previous order. Another example is Pump \& Dump, where the manipulator makes a profit by a quick flipping of long-held holding of shares at the manipulated price.

Diaz et al.  \cite{diaz2011analysis} show that it is possible to discover manipulation patterns from hourly transactional data using knowledge discovery techniques. They adopt an open box approach that uses heterogeneous sources such as news sources, financial ratios, variables, etc., to describe trades at an intraday (i.e., hourly) level. The dataset used in this research is based on the stock market manipulation cases pursued by the SEC, and is also the dataset that we will use in our work.

In the work of Jia \cite{jia2016investigation}, the effectiveness of LSTM is explored for stock price prediction. Though this research has no connection with any type of fraud or manipulation detection, it can still be categorized as a manipulation detection process.

In the work of Song et al. \cite{song2012coupled}, the authors propose a general Coupled Behaviors Analysis (CBA) framework for detecting group-based market manipulation by capturing more comprehensive couplings. Groups of manipulators collaborate with each other to manipulate the stock price, which is a big challenge for stock market surveillance. Their approach performs better than the previous benchmark Coupled Hidden Markov Model (CHMM), and the main reason for this performance was proper domain knowledge among stocks.

In the work of Zaki et al. \cite{ali2014analyzing}, they propose a linguistic-based text mining approach to demonstrate the process of extracting financial concepts (e.g., parties involved, jurisdiction, financial gain by the fraudster) from SEC Litigation Releases (LR). They use a financial ontology to capture financial fraud concepts from the SEC litigation releases. This helps financial analysts to understand different manipulation patterns from the prosecuted cases. However, their work has a limitation on the number of litigation cases studies. Usually, the more litigation cases that can be studied, the less skewed the outcome. On the other hand, we are using a tree-based approach, which will aid analysts by filtering the number of cases they need to examine. Moreover, our proposed approach helps to mine potential illegal insider trading in advance (i.e., prediction). 

In the work of  Mantere et al. \cite{mantere2013stock}, they present a hypothetical case study of stock market manipulation using cyberattacks together with false information disseminated through social media so that the false news influences the market in a favorable way. 

So far we have found a variety of approaches for stock market manipulation fraud detection. Some approaches are based on peer group analysis, some are based on contextual anomaly detection, some are based on linguistic-based text mining, and some use financial ontology. There is very little research in the area of detecting illegal insider trading in the stock market. The work of \cite{ahern2017information} is based on insider trading but the research just analyzes insider trading cases. To the best of our knowledge, what we are proposing is a more comprehensive approach detecting illegal insider trading. To the best of our knowledge, due to the novelty of this work, we will be unable to benchmark our work against other approaches.

\section{Methodology}\label{methodology}

\begin{figure}[t]
\centering
\includegraphics[scale=.54]{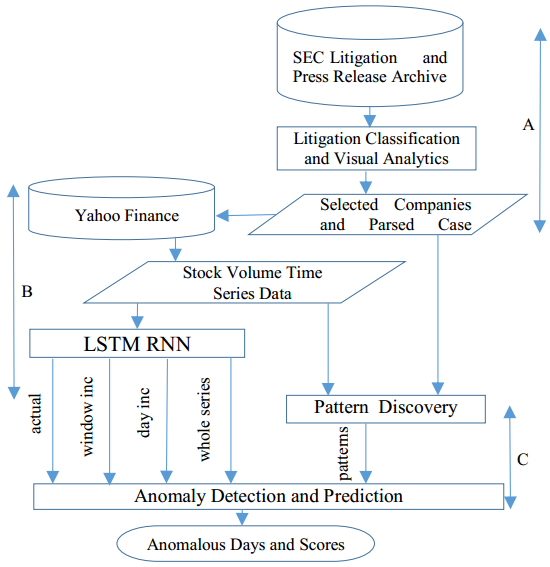}
\caption{ Flowchart of the proposed technique. }
\label{fig:Flowchart}
\end{figure}

\subsection{Proposed Technique}
Figure \ref{fig:Flowchart} is the flowchart of the technique that we implemented. In the beginning, we targeted companies (Figure \ref{fig:Flowchart}, labeled as A) that were identified by our tree-based visualizations (to be discussed later in this paper) and also identified as prominent illegal insider trading cases by the SEC. Then, we collected historical stock volume data from Yahoo Finance \cite{YahooFin39:online} for the targeted companies. The start date of the collected stock volume data is three months earlier than the date fetched from the parsed case and press release, when the insider learns the non-public information. And the end date of the collected stock volume data is the date when we performed these tests. We collected all day-wise stock transaction volume for the targeted companies within these start and end dates (Figure \ref{fig:Flowchart}, labeled as B).

In pattern discovery stage (Figure \ref{fig:Flowchart}, labeled as C), we selected the windows (a window is a collection of transaction volumes for 50 consecutive days) when the insider learns the nonpublic information and did transactions according to that. We found that, in most of the cases, the insiders do the transaction within a very short time frame (one to two days) of the learned nonpublic information. So, we selected 50 days of transaction volume for a targeted company in such a way that the date when the insider learns the non-public information is in the middle. Some of the companies face illegal insider trading multiple times (e.g., Wells Fargo), so we get multiple patterns (window) from those companies. By this way, we generated all illegal insider trading patterns, which are ground truth for our anomaly detection algorithm ANOMALOUS.  

Our next task is to predict stock volume (Figure \ref{fig:Flowchart}, labeled as B and C) for the targeted companies for a certain period and pass the discovered patterns as a sliding window over it to see the similarity.   
So, for the selected companies, we apply LSTM RNN to the historical stock transaction volume data using three different approaches: 

\begin{enumerate}
    \item predicting a future window of transaction volume (e.g., 50 days), based on the previous window of transaction volume, 
    \item predicting transaction volume of a day ahead based on the previous window of transaction volume, and 
    \item predicting a future window of transaction volume, based on all previous transaction volume data.
\end{enumerate}

It is necessary to mention that, we define a window as a collection of N consecutive days. In this work, we use a window of size 50, which is the same window size used by Jia et al. \cite{jia2016investigation}, albeit, the purpose of their work is to predict stock prices using LSTM RNN, whereas, we are using transaction volume data to predict stock market transaction volume towards the prediction of illegal insider trading by our proposed ANOMALOUS algorithm. By analyzing data, we realize that in the case of illegal insider trading, the transaction volume is a better candidate  to consider for the algorithm as the price of an S share might be much less or more than S, but the transaction volume for S share is going to be exactly S.

In addition, the output from LSTM RNN consists of four types of discrete signals (time series): the actual data (transaction volume), window wise predicted signal (transaction volume for next window), day wise predicted signal (transaction volume for next day), and signal (transaction volume for next window) based on the whole historical data. For anomaly detection, we used normalized cross-correlation that can measure the similarities between given discrete signals with different time lags. We represent transaction volume of consecutive days as a discrete signal. The generated four types of discrete signals and previously generated illegal insider trading patterns are passed through our proposed anomaly detection and prediction algorithm (ANOMALOUS).We named it ANOMALOUS as it helps to find Anomalous Signals (time series).  

The parameters for the ANOMALOUS algorithm are C, M, W, P, and Data. C defines the selected companies, M defines the specific application (i.e., one of three techniques) of LSTM RNN, W defines the number of windows, P is the extracted pattern (i.e., ground truth) for illegal insider cases, and Data is a multi-dimensional array representing the predicted stock transaction volume data by the LSTM RNN. The ANOMALOUS algorithm is presented below. For each company c in C, and each specific way m in M applications of LSTM RNN, and each window w in windows W, signals are compared with illegal insider trading patterns or anomalous patterns to discover how similar they are by segmenting the whole time series into windows and further dividing the windows into days. The predicted signals (output of LSTM RNN) are compared with both actual signals (actual  stock transaction volume from historical data) and  fraudulent/anomalous signals/patterns (discovered illegal insider trading pattern) using Normalized Cross Correlation (NCC) which actually normalize the signals into the same scale and then compare signals to see how strongly they are correlated. In addition, the NCC measures the correlation of two discrete signals by considering different day lags. Here day lags refer to the knowledge that one of the discrete signals has some missing value at the beginning/end of the signal, as compared to the other signals. In other words, day lags means the signals may not be exactly vertically aligned with one another as there may be some gaps in days. The three types of prediction techniques using LSTM RNN and the detail of NCC are described in Section \ref{experimentalresult}. Here is our proposed anomaly detection algorithm that visualizes and matches the predicted and actual time series with anomalous time series patterns.

\begin{figure}[h]
\centering
\includegraphics[scale=.6]{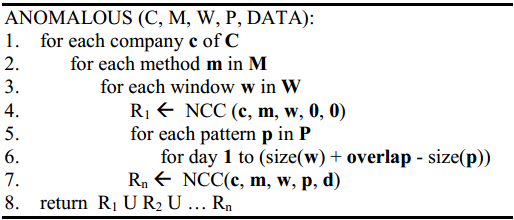}
\label{fig:Anomalous}
\end{figure}

\subsection{Data}
The SEC releases all litigation’s brought by the commission in federal court on their website \cite{SECgovLi22:online}. This includes all types of stock market manipulation charges. In addition, the SEC and FBI also publish some of the prominent illegal insider trading cases and related press releases on their website \cite{SECEnfor61:online,Financia28:online}. These are the primary source of the data for our research. We have written a Python-based web crawler that crawls through litigation related press releases archive starting from the beginning year 1996 to the current year 2018 and downloads and stores them as plain text. In total, we found 7988 cases, and out of that, 605 cases have the keyword ``insider'' in the title of the release, 1142 cases have the keyword in the body of the release, 1222 cases have the keyword in either the title or body of the release, and 525 cases have the keyword in both the title and body of the release. So, roughly around 15\% of the total cases were illegal insider trading-related charges and the remaining were other kinds of fraud-related charges. Based on the keyword ``insider'', we labeled all cases as insider cases or non-insider cases. And then using that data, we build a classifier for classifying available illegal insider trading related documents, enabling us to classify future data (e.g., social network posts). We also generated tree-based visualizations for analysts to understand how a case falls into the illegal insider trading category and the correlation between events and actions. From the visualizations of actual cases, we are able to drill down into the litigation releases and retrieve relevant attributes (e.g., company name, illegal insider name, the date when the insider learns the private information, the date when it was public, and the volume of illegal gains, etc.) (It should be noted that currently this process is manual, and an automated solution will be investigated as part of future work.).

Another source of data for this work is the historical stock transaction volume (i.e., selected day-wise, transaction volume data for selected companies) data from Yahoo Finance \cite{YahooFin39:online}. We collected thousands (14,842) of time series data for the past few years (all that was available publicly at that time) for the ten different companies for the experiments. Out of the ten companies, nine involved in historical illegal trading cases: Wells Fargo, BP plc, GTx, Oracle Corporation, American Semiconductor Corporation, Spectrum Pharmaceuticals, Allscripts Healthcare Solutions Inc, Herbalife Ltd, Evercore Inc. The 10\textsuperscript{th} company is Google's parent company Alphabet Inc. which was created in 2015 and there are no reported illegal insider trading cases so far. All publicly available time series stock volume data for Alphabet is selected and used for validating the anomalous patterns – to see whether any window from Alphabet Inc. matches with any of the anomalous patterns that we explored in our experiments.  We used the transaction volume data feature as we found that transaction volume fluctuates more than any other available features. Furthermore, we transformed the transaction volume data into a percentage of change with respect to the starting day of the window or sequence for better interpretability of the result. For example, if first three days of a window (of size 50 days) has transaction volumes of 100, 120 and 115, then the actual data that is fed in to the neural network is  0, .20, and .15 (i.e., the percentage change with respect to 100, which is the transaction volume of day 1 for the  current window). 

\section{Experimental Results}\label{experimentalresult}
The following sections represent the experimental steps corresponding to our proposed architecture presented in Figure \ref{fig:Flowchart} along with results.

\subsection{Litigation Classification and Visual Analytics}
Our first experiment is preprocessing a huge number of unstructured text data using combined Natural Language Processing (NLP) and Decision Tree-based approach. And its main purpose is to aid an analyst's  understanding of the data. Our first attempt was to finding discriminant features from the text that helps to distinguish insider trading cases from the case archive. It is necessary to point out that, initially we label all cases as insider cases that have the insider keyword in the title or body of the case. We tokenize all the cases and make the feature vector (a matrix where the columns contain the features/tokens and the rows contain the case wise token/feature frequencies). Later we run the Extra Trees algorithm to rank the features. The Extra Trees (ET) algorithm, also known by the term Extremely Randomized Trees, is an ensemble tree-based approach where the randomness goes further compared to the Random Forest algorithm \cite{geurts2006extremely}. Here the splitting attribute is also chosen in an extremely random manner in terms of both variable index and splitting value so as to randomize the tree, whose structure has no relation with the learning samples \cite{islam2018credit}.

Furthermore, the Extremely Randomized Trees algorithm has shown positive results in some fraud/anomaly detection research \cite{islam2018credit,islam2018efficient} using public datasets. We set the biased parameter of the ET algorithm to maximum so that it gives more priority to correctly classifying the insider cases as opposed to the non-insider class. Some of the discriminant features that we found from it are as follows: insider, insiders, friend, classmate, non-public, linked, etc. We also run TF-IDF to find the features that are uncommonly common in the cases. Also, given that calculating TF-IDF requires a substantial amount of memory for the size of the data set. We used the Apache Spark tool where we can distribute the work to multiple commodity machines and gather the result back to the master node for returning the accumulated result. This tool also allows us to create a data frame (data container) whereby if the data size is more than the capacity of RAM, disk is used to hold the remaining data. Thus, we can use a single commodity machine (with 12GB RAM and a core i7 processor) for this experiment when the processing time is not a major concern. But this is easily scalable to multiple nodes with the help of Apache Spark when the feature vector for TF-IDF is large and/or faster processing is a major concern. Almost all of the top results from the TF-IDF algorithm only returned human names as being the most useful feature, rather than important illegal insider trading keywords that were found to be the most useful features by the Extra Trees algorithm. Thus, we only kept the features identified by the ET algorithms, discarding all other features that have a feature rank of zero (i.e., no contribution in the classification process\textemdash zero information gain). After that, we made a new feature vector with the filtered features (i.e., reduces features) which were smaller than before and easily usable on a commodity laptop.
\begin{figure}[t]
\centering
\includegraphics[scale=.43]{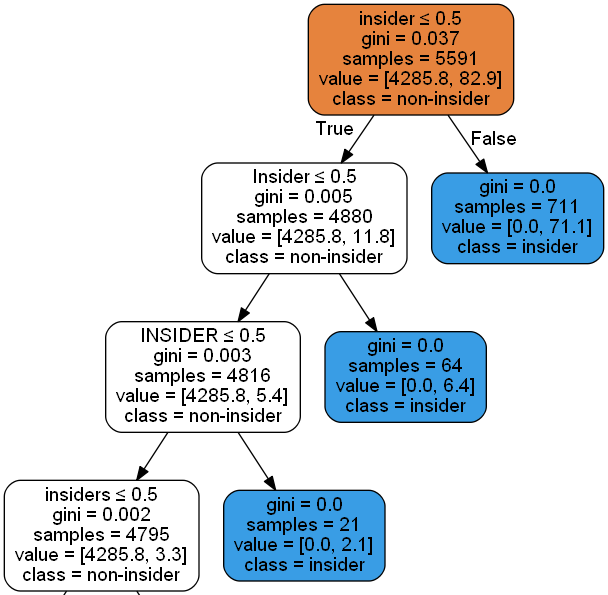}
\caption{ A portion from the top of the decision tree showing the decision path.}
\label{fig:tree1}
\end{figure}
\begin{figure}[t]
\centering
\includegraphics[scale=.41]{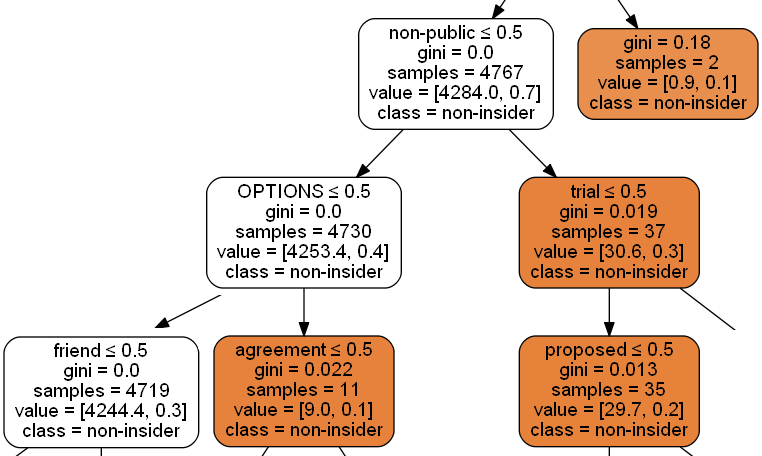}
\caption{ A portion from the middle of the decision tree showing the decision path.}
\label{fig:tree2}
\end{figure}
We use the decision tree algorithm on the processed data (i.e., reduced feature vector) from litigation cases and visualize the important textual features and their relations in terms of nodes and edges of the tree. This helps analyst to understand the importance of the insider cases, parties involved, etc.  The developed model can also be used to classify new unlabeled cases or news article from other sources for insider vs other case classification. We hope to include more data sources in our future work on this. 
Figure \ref{fig:tree1} and Figure \ref{fig:tree2} show portions of the tree generated by running the decision tree algorithm on the reduced features. The features (1526 features) were selected using Extra Trees Algorithms. Examining the tree, we can see that the features such as insider and its variants (e.g., insiders) of non-public, friend, proposed, trial, etc., play a key role in decision making. These features can be found in the upper section of the ranked features list. From the analysis, we also found that in most of the cases, the information was leaked by friends, colleagues, and relatives in their social or work life. The blue color nodes represent the insider class, the brown color nodes represent the non-insider class, and the white color node (can be insider or non-insider class) represents the decision path for a particular sample. In short, the decision tree acts as a filter (or funnel) for narrowing down the possible number of litigation cases that would need to be considered by a fraud analyst. In addition, from these discovered cases, illegal insider trading patterns can be generated for classification of future cases. The top-ranked features from TF-IDF are mostly human names: Donovan, Abe, Patel, Steffes, and Keith. However, unfortunately we do not have any more information on these individuals. In the future, we want to investigate the incorporation of other data sources, like social media, that would provide more insight as to what transpired. 

\subsection{Predicting Company Wise Stock Transaction Volume}
For predicting stock volume, we used the deep learning technique LSTM RNN with the help of Tensorflow in the back-end and Keras as a wrapper. The input layer consists of 50 neurons which correspond to the window size. In other words, each neuron is fed with the stock volume for a day. This input layer is then fed into an LSTM layer of 50 neurons which is connected to another LSTM layer of 100 neurons. The last LSTM layer is connected to a fully connected normal layer of 1 neuron with a linear activation function. This activation function is used to give the prediction for the next time step. Other configurations for this experiment are: dropout = .2, loss = mse, optimize= rmsprop, epoch = 1, and batch size = 512. The rmsprop optimizer divides the learning rate for a weight by using a running average (i.e., average changes/updates continually as new data points collected) of recent gradients for that weight.
\begin{figure}[h]
\centering
\includegraphics[scale=.42]{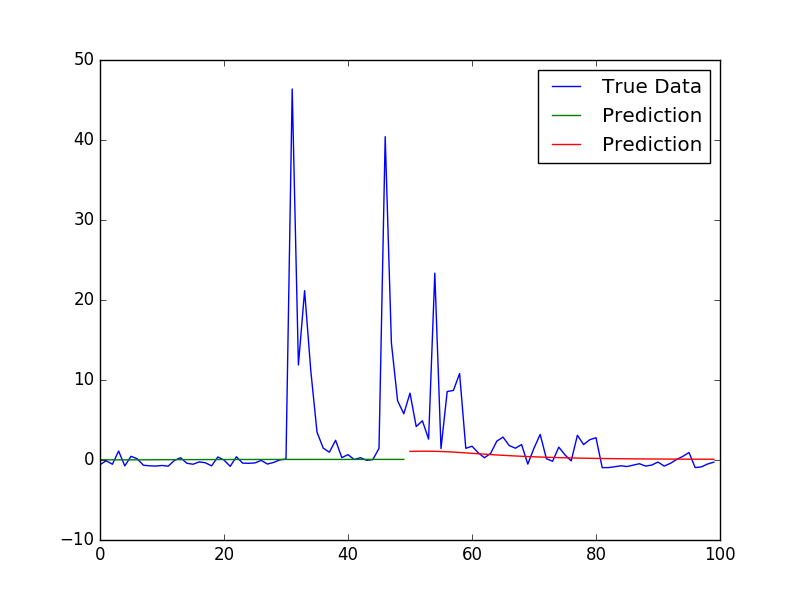}
\caption{ Window-based prediction.}
\label{fig:window}
\end{figure}
To recall, we applied LSTM RNN in three different ways. Figure \ref{fig:window} is a visualization of the window based prediction method where for the prediction of window w, only window w-1 is used. The x-axis represents the time series. Here we have two-time series each consisting of 50 days. And the Y-axis represents the percentage of change in stock transaction volume from the starting day of the window which means that the Y value of the starting day of any window will be zero.
Furthermore, Figure \ref{fig:day} is the visualization of day based prediction where all previous days in the window are considered for making a prediction for the next day, and the knowledge base is updated after each day with the true observation. Figure \ref{fig:entire} is the visualization of a sequence-based prediction where entire history (all windows) is considered while making a prediction.
\begin{figure}[h]
\centering
\includegraphics[scale=.4]{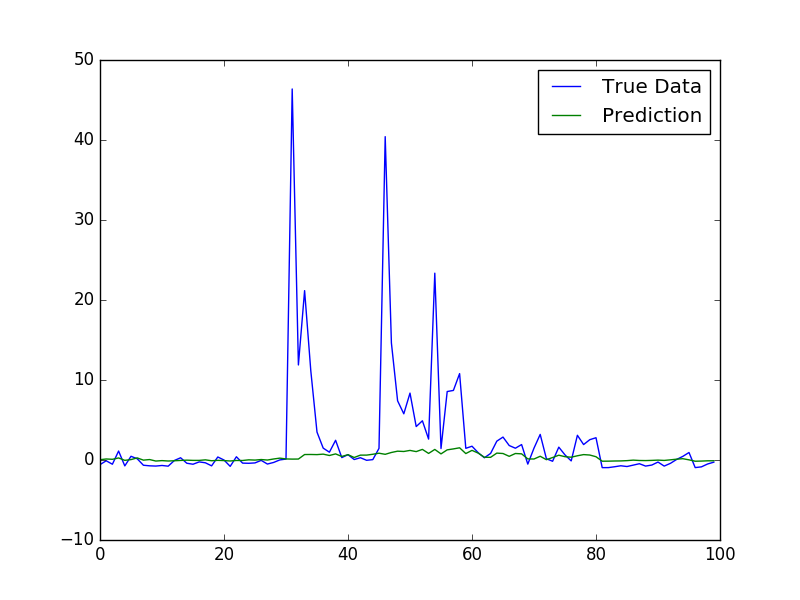}
\caption{ Day based prediction.}
\label{fig:day}
\end{figure}
\begin{figure}[h]
\centering
\includegraphics[scale=.4]{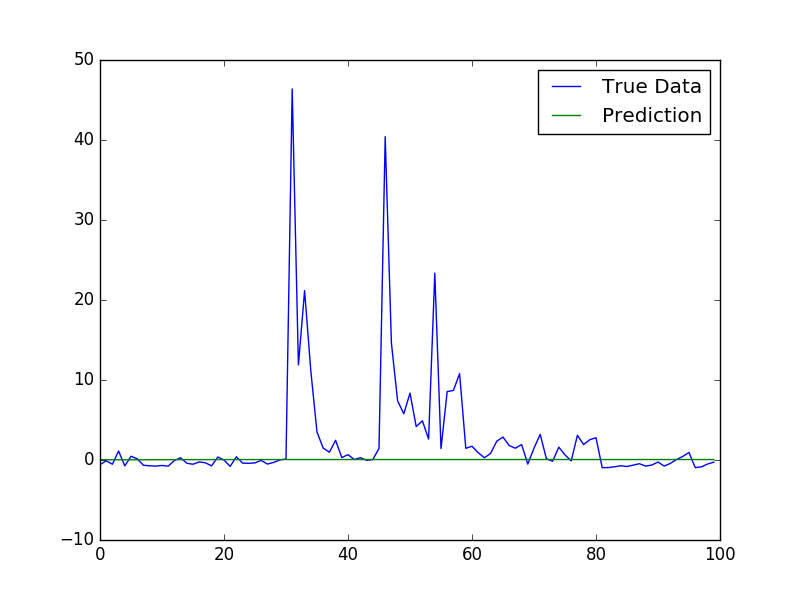}
\caption{ Whole history-based prediction.}
\label{fig:entire}
\end{figure}
After the stock transaction volume prediction has been completed, we then forward the predicted data (i.e., output from LSTM RNN) to ANOMALOUS algorithm for anomaly detection as a multidimensional array. The output from LSTM RNN is three discrete signals for the three different ways we applied the LSTM RNN.

\subsection{Anomaly Detection and Prediction}
Our last experiment is to apply our proposed ANOMALOUS algorithm for comparing different discrete signals. In this stage, we wrote a Matlab script to implement the algorithm. For the comparison of discrete signals, we used the Normalized Cross Correlation (NCC). The Correlation indicates the similarity of signals and is widely used in various applications (e.g., speech recognition) for signal comparisons. In this setting, we are using a normalized version of Matlab’s cross-correlation rather than the standard version of cross-correlation because we need to know the similarity   between signals of different time series (i.e., different minimum and  maximum values) in the same scale (-1 to 1). The formula for the cross-correlation is: 
\begin{equation} \label{eq:formula1}
 Corr_{x,y} = \sum_{n=0}^{N-1} x[n]y[n] 
\end{equation}

For our experiments, x and y are the vectors of time series data consisting of the stock transaction volume of a defined consecutive number of days (i.e., the window), where N represents the number of days in the series. Thus, the cross-correlation (Corr) is simply the sum of the scalar multiple of corresponding elements of two signals or time series, where the higher the value the more they are correlated or similar. But the companies in comparison have transaction volume of different ranges (e.g., maximum transaction volume for Wells Fargo is 254,575,800 whereas for Evercore Inc. it is only 4,345,100).   Since the signals being compared have different energy levels or aptitudes values, we applied a normalized version of cross-correlation (NCC) that converts those into the same scale first and then calculates the correlation. The Normalized Cross Correlation (NCC) is formulated as follows:
\begin{equation} \label{eq:formula2}
Corr{\text -}norm_{x,y} = \frac{\sum_{n=0}^{N-1} x[n]y[n]}{\sqrt(\sum_{n=0}^{N-1} x[n]x[n] \sum_{n=0}^{N-1} y[n]y[n])}
\end{equation}
Here the nominator part is exactly the same as before (Formula \ref{eq:formula1}) which is scaled using the factor of the energy level of both participating signals for the purpose of normalization. In Matlab, it is available as a built-in feature and named as xcorr. To apply the normalize option, we need to pass the parameter coeff along with the parameters of the signal.

\begin{figure}[h]
\centering
\includegraphics[scale=.56]{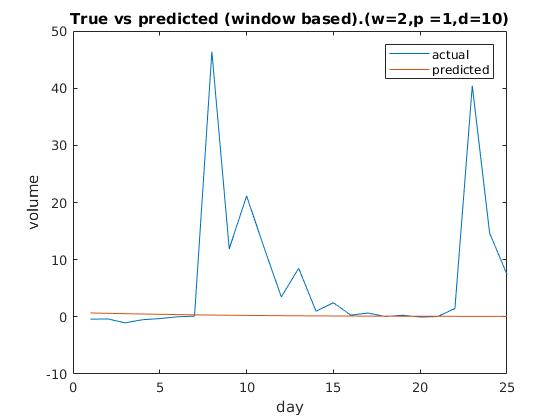}
\caption{ An anomalous time series (window-based).}
\label{fig:anomalous-window-based}
\end{figure}

\begin{figure}[h]
\centering
\includegraphics[scale=.56]{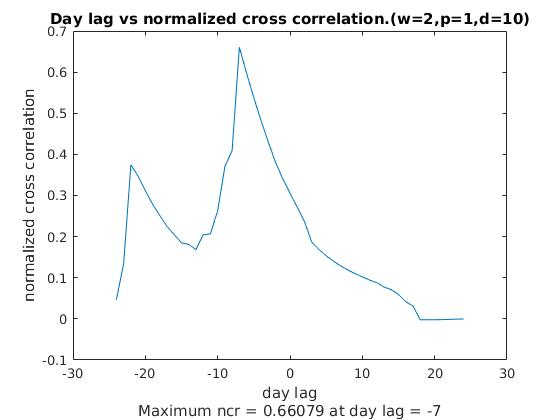}
\caption{ NCR (window-based).}
\label{fig:ncr-window-based}
\end{figure}

Figure \ref{fig:ncr-window-based} tells us that the two time series from Figure \ref{fig:anomalous-window-based} are most similar at day lag -7 (position of the spike) with a correlation value of .661. Since the maximum value of correlation is 1 (when they are exactly the same), this value of .667 tells us that they are not that similar. If we want to get the actual day when the value of NCR is the highest in the chart (e.g., Figure \ref{fig:anomalous-window-based}), we can use our formulated equation (3) by observing the data:

\begin{equation} \label{eq:formula3}
day = window\_size + (w * window\_size) + d
\end{equation}
For our experiment, the window size = 50. The value of w and d are mentioned on the top of the figure (Figures 7-12), p tells the pattern number in comparison.  For example, the calculation of the day where the NCR is highest from Figure \ref{fig:anomalous-window-based} using equation \ref{eq:formula3} is as follows:
day =  50 + (2 * 50) + 10 

\begin{figure}[h]
\centering
\includegraphics[scale=.56]{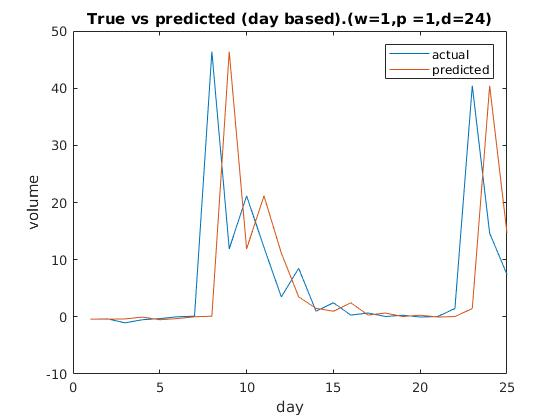}
\caption{ An anomalous time series (day based).}
\label{fig:anomalous-day-based}
\end{figure}

\begin{figure}[h]
\centering
\includegraphics[scale=.56]{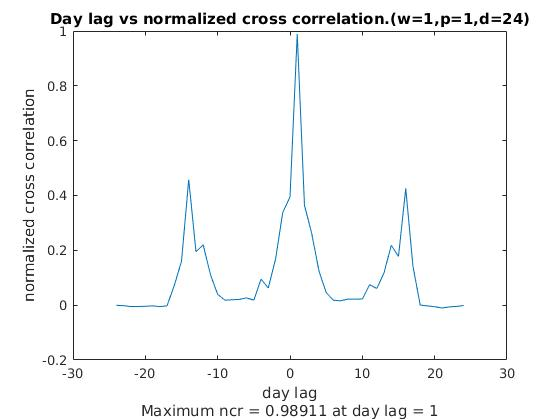}
\caption{ NCR (day based).}
\label{fig:ncr-day-based}
\end{figure}
= 160. So it is the 160\textsuperscript{th} day in respect to the beginning of the entire time series. 

From Figure \ref{fig:anomalous-day-based}, it is clearly visible that the predicted result is almost similar to the actual result. Figure \ref{fig:ncr-day-based} is the correlation between the above signals in Figure \ref{fig:anomalous-day-based}, which tells us that the time series or signals are almost similar at day lag 1 with a correlation value of .989. Here day lag 1 means if we shift the actual signal right by 1 day then it will give us a correlation value of .989 that we got.

\begin{figure}[h]
\centering
\includegraphics[scale=.56]{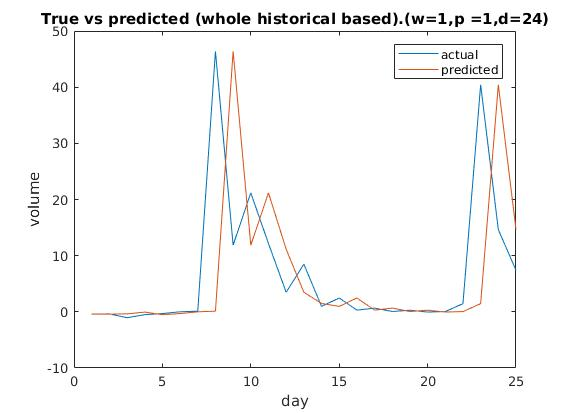}
\caption{ An anomalous time series (whole history based).}
\label{fig:anomalous-whole-history-based)}
\end{figure}

\begin{figure}[h]
\centering
\includegraphics[scale=.56]{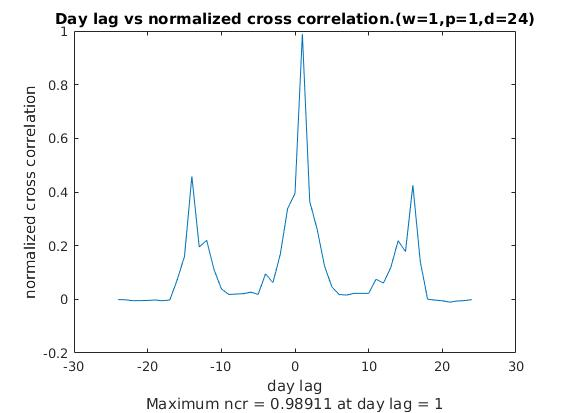}
\caption{ NCR (whole history based).}
\label{fig:ncr-whole-history-based)}
\end{figure}

Figure \ref{fig:anomalous-whole-history-based)} and Figure \ref{fig:ncr-whole-history-based)} are the visualizations and correlations of the entire time series based prediction.  However, while the results seem promising, we are only using a small set of windows.  In the future, we plan on experimenting with longer runs where the number of windows increases with more data points \textemdash that will also help us to better understand the gap between some of the peaks of the actual and predicted data. Because it is difficult to predict time series data with traditional prominent machine learning approaches (e.g., SVM), these initial experiments indicate potentially good results can be realized using a deep learning based technique like LSTM RNN – especially within our proposed framework where we can control how much information we want to use for better predictions (e.g., for window based prediction, just the previous window is used for predicting next window). 

\begin{table}[]
\caption{Pattern wise hit}
\renewcommand{\arraystretch}{1.2}
\begin{tabular}{|| c c c c c c ||} \hline
Pattern & \% Companies & Total hit & Window & Day & History\\ 
\hline \hline
1       & 100\%                       & 74        & 8            & 66        & 0                       \\
2       & 100\%                       & 128       & 116          & 8         & 4                       \\
3       & 28.57\%                     & 5         & 0            & 5         & 0                       \\
4       & 57.14\%                     & 4         & 0            & 4         & 0                       \\
5       & 71.42\%                     & 76        & 11           & 65        & 0                       \\
6       & 28.57\%                     & 10        & 1            & 9         & 0                       \\
7       & 14.29\%                     & 1         & 0            & 1         & 0                       \\
8       & 14.29\%                     & 4         & 0            & 4         & 0                       \\
9       & 14.29\%                     & 1         & 0            & 1         & 0                       \\
10      & 14.29\%                     & 1         & 0            & 1         & 0                       \\
11      & 14.29\%                     & 1         & 0            & 1         & 0  \\     \hline 
\end{tabular}

\label{table:1}
\end{table}

\begin{figure}[h]
\centering
\includegraphics[scale=.6]{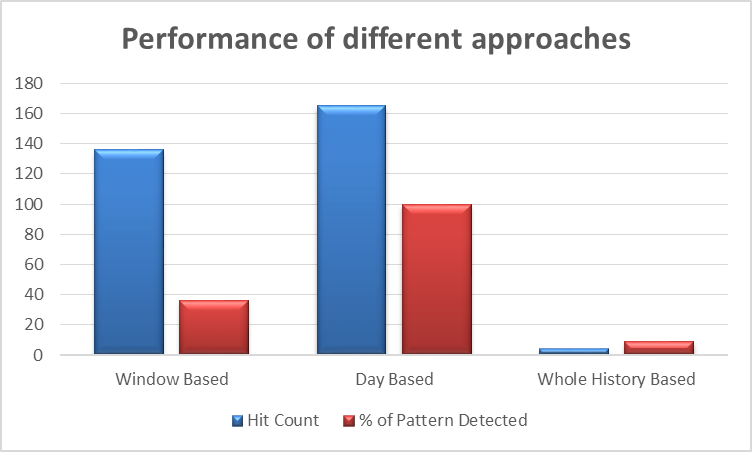}
\caption{ Performance of different approaches.}
\label{fig:performance}
\end{figure}

Finally, Figure \ref{fig:performance} and Table \ref{table:1} show the comparison of three variants of LSTM RNN that we applied in our experiments. First, the day-based prediction approach outperforms other approaches as it is capable of detecting all anomalous patterns with the highest hit count. Day-based prediction performs better because it uses the true history of all previous days for that window. But the limitation of this approach is that the number of days that it can predict in advance is only one. On the other hand, the window-based prediction approach predicts a whole window (50 days) in advance based on the previous window, so it performs a little worse than the day-based prediction. Finally, the entire history-based prediction approach performs the worst as it does the prediction using the entire historical data set, which is impractical for time series prediction. Furthermore, Table \ref{table:1} shows pattern-wise results (grouped by methods) for discovering the anomalous patterns. The first two patterns (pattern 1 and 2) are detected in all of the companies multiple times in the whole time series. But the actual illegal insider trading case was just for two different companies in two different time spans. So our approach identifies several suspicious time series whose actual nature may not only be illegal insider trading, but could consist of other potential financial fraud situations.  However, there may also be false positives.  One possibility to mitigating the false positives is to increase the NCR correlation threshold (currently set to .80), but that will also result in losing some of the anomalous patterns.

To validate how good indicator are those anomalous patterns, we also run our experiment on Google's parent company Alphabet Inc. which was created in 2015 and there are no reported illegal insider trading cases so far. We find that with a correlation coefficient of .94 and above there is no available anomalous
pattern. When we reduce the value below to .94, several time-spans are marked as similar with the anomalous patterns (illegal insider trading time series) that we have identified from the historical data as actual insider trading cases.

Predicting illegal insider trading correctly is almost a nightmare like predicting stock market price correctly. What we can do is make an educated guess using different scientific approaches. If a time series is identified as suspicious then it can be a result of something legal or illegal. If it is illegal then it can be any type of fraud activities. As we trained the model using some true illegal insider trading pattern so there is a good chance that our
model's prediction can be an actual illegal insider trading. Taking measures proactively using our approach can mitigate damage in the market before the official news of the illegal insider trading is released. After all, the insider may be aware of familiar insider trading patterns and might want to evade the detection. This is a known limitation of our dataset, and in the future we will try to integrate more data sources (e.g., social media) to tackle this issue.

\section{Conclusion}
In this work, we presented how to predict and detect illegal insider trading proactively (before the official news is released) by analyzing heterogeneous sources of structured and unstructured data. We also showed an effective way (i.e., tree-based visualizations) to intuitively represent the unstructured data to the analyst for better understanding and a possible reduction of the search space. We targeted companies that were identified by our tree-based visualizations and also identified as prominent illegal-insider trading cases by the SEC. For the targeted companies, from the parsed litigation and historical stock data, we discovered the illegal insider trading patterns (anomalous pattern). Later we used LSTM RNN to predict the stock transaction volume using three approaches. After that, we used our proposed algorithm to see whether any of the time spans (actual or predicted) match with the discovered anomalous patterns. We found that our algorithm has a good success rate in detecting an anomalous pattern from both the affected and non-affected companies. One limitation of this research is that we were not able to compare our result with any similar state-of-the-art work because our work is the first work (to best of our knowledge) in detecting specifically illegal insider trading from true litigation cases.

A future direction of this research is trying the prediction and detection part with a larger number of companies to see its effectiveness. In addition, test and compare the prediction part of this work with some other algorithms (e.g., GAM, ARIMA) that work well with time series prediction.

\section*{Acknowledgment}
Our thanks to Jakob Aungiers for the Python-based implementation of stock price prediction using LSTM RNN. In the first part of our experiment, his code was a great help to us. We used his code as a template, tuned some parameters, and modified it to adapt to our experiments.
\bibliographystyle{IEEEtran}
\bibliography{IEEEabrv,bib/references}
\end{document}